\documentclass[prd,preprintnumbers,twocolumn,amsmath,amssymb,nofootinbib,showpacs]{revtex4}

\usepackage{epsfig,amssymb,amsfonts,verbatim}

\def\bfl{\begin{flushleft}}
\def\efl{\end{flushleft}}
\def\bfr{\begin{flushright}}
\def\efr{\end{flushright}}
\def\bc{\begin{center}}
\def\ec{\end{center}}
\def\be{\begin{equation}}
\def\ee{\end{equation}}
\def\ba{\begin{eqnarray}}
\def\ea{\end{eqnarray}}
\def\baa#1{\begin{array}{#1}}
\def\eaa{\end{array}}
\def\bw{\begin{widetext}}
\def\ew{\end{widetext}}

\def\lb#1{\label{#1}}

\def\drm{\text{d}}

\def\sfv{V}
\def\Eq{eq. }
\def\Eqs{eqs. }
\def\Rf{ref. }
\def\Rfs{refs. }

\def\ap{{A}}

\def\tp{{\tilde p}}

\newtheorem{Def}{Def.}[section]

\newtheorem{Prp}[Def]{Proposition}

\def\k{\kappa}

\def\s{\sigma}

\begin{document}

\preprint{Phys. Lett. B\textbf{638} (2006) 89-93}
\preprint{hep-th/0601221}

\title{
Why do we live in a 4D world: Can cosmology, black holes and branes give an answer?
}

\author{Konstantin G. Zloshchastiev}

\affiliation{Instituto de Ciencias Nucleares,
Universidad Nacional Aut\'onoma de M\'exico, A.P. 70-543,
 M\'exico D.F. 04510, M\'exico}





\begin{abstract}
We  derive the general form of the cosmological scalar field potential
which is compatible both with the existence of  black holes and
p-branes related to string/M theory
and with  multidimensional  inflationary cosmology.
It is shown that the scalar potential alters non-trivially from dimension to dimension
yet always obeys one single equation where the
number of spacetime dimensions is a free parameter.
Using this equation
we formulate an eigenvalue problem for the dimensionality parameter.
It turns out that in the low-energy regime of
sub-Planckian values of the inflaton field,
i.e., when the Universe has cooled and expanded sufficiently enough, the
value four arises as the largest admissible (eigen)value of this parameter.

\end{abstract}

\pacs{11.25.-w, 98.80.Cq, 04.70.Bw}
\maketitle




Modern trends in high-energy physics such as
string/M theory suggest that the
unification of all fundamental interactions is impossible without increasing
the number of actual spacetime dimensions, $D$, from $4$ to $10$ or even $11$.
It seems that only in higher dimensions one can reconcile general relativity
with quantum mechanics and construct a self-consistent theory.
Nevertheless, despite the continuously growing popularity and successes of higher-dimensional models one should not
forget one of the fundamental problems of string/M theory: to explain why
at low energies
our world is
effectively four-dimensional  to such a high degree of precision.

Recent efforts to dynamically derive or explain 4D spacetime have been made in the matrix
formulation of type IIB superstrings
but no definite conclusion has been made so far
\cite{nishimura3,nishimura4,bielefeld}.
Also, attempts have been made in alternative theories of quantum gravity, such as
Causal Quantum Gravity \cite{Ambjorn:2004qm} but that theory has too little in common
with M theory.

Other attempts were also made in several multidimensional models
imposing one or another predefined way of splitting a higher-dimensional manifold
into the external space (the observable Universe) and an internal one.
For instance, working in 11D supergravity quantum cosmology if
one assumes the seed instanton of the Universe to be a product of two spheres,
then it can be proven that the external spacetime must be 4D \cite{Wu:2001id}.
Although, in such models the problem was not entirely solved but rather reduced to the well-known
problem of which way of dimensional reduction the Universe uses to descend from 11D to 4D.
This is still far from being answered.

Thus, the problem of formation of the Universe's dimensionality still persists.
Is it possible to solve it in a way consistent with M theory ideas?
In this paper we will try to get an affirmative answer.

In fact, as compared to all other parameters which appear in field theory, the parameter $D$ has a number
of distinctive
features which, on one hand, create difficulties with solving the above-mentioned problem but,
on the other hand,  hint about possible clues.
First, we know that any field theory in spacetime is defined by virtue of the action functional,
$
S [\psi (x)] = \int L (\psi_i, \partial \psi_i) d^D x
$,
where $\psi_i$ is a set of the fields built into the theory, $L$ is  the Lagrangian function.
Here the spacetime dimensionality $D$ is involved both in
the integration measure of the functional, $d^D x$, and in the notion of distance
between events in spacetime, thus
it is the parameter which must be fixed \textit{before} one starts specifying the Lagrangian itself.
Therefore, $D$ can not be directly derived from a theory which acts in spacetime.
Second, being an integer rather than a continuous  parameter, $D$ hints about the
possibility that it could be
a discrete eigenvalue of
a certain dynamical theory (which acts not in spacetime, according to the aforesaid).

Below we  show that both suppositions are indeed very close to the truth.
We demonstrate that the values of the parameter $D$ are dynamically derivable
from a differential equation defined in a certain space (called the coupling space)
whereas the properties of the latter
are determined by  black hole physics closely related to M theory, supersymmetry (SUSY) and p-branes.

We start by recalling that the global dynamics of the Universe is determined by long-range
fields.
From the Standard Model we know that Yang-Mills and fermionic fields are short-range  thus
here we can restrict ourselves to the Abelian sector.
The latter includes gravity described by the Lorentzian metric tensor $g_{M N}$ (capital
Latin indices run from $1$ to $D$), real scalar field $\phi$ (often called also the
\textit{inflaton} as it has been widely used in cosmology in  scalar-driven inflationary models) and
several gauge fields described by antisymmetric tensors of different ranks.
Thus, our action functional must contain at least the graviton and the inflaton.
To this we add also a couple of gauge fields: the electromagnetic one described
by the antisymmetric tensor of second rank, $F_{M N}$, and the one of
rank $\tilde p$, with the components $F_{(\tilde p) M ... N}$.
The gauge fields are added just to check whether they affect the problem we are going to study.
The simplest effective action one can thus write is (we work in the Einstein frame):
\be
 S =
 \int d^D x \sqrt{- g} \,
\biggl[
      R - \frac{1}{2} (\partial \phi)^2
      + \Xi (\phi) F^2 + \Psi(\phi) F_{(\tilde p)}^2 -V (\phi)
\biggr],
\lb{eEMD}
\ee
where
$ (\partial \phi)^2 = \partial_{M} \phi \, \partial^{M} \phi$,
$F^2 = F_{M N} F^{M N}$, $F_{(\tilde p)}^2 = F_{(\tilde p) M .. N} F_{\!(\tilde p)}^{\ \ M .. N}$,
and
$g=\det{|g_{M N}|}$
and
$R$ is the scalar curvature constructed out of $g_{M N}$.
By this action we mean an \textit{ensemble} of  theories  which are equivalent with respect
to their dynamical content
but
act in different $D$.
Then we can regard $D$ as a free parameter from now on.
For further, it will be convenient to assume
$D \equiv \tilde p + 2$ and work  in terms of the parameter $\tilde p$.

The explicit values of
the Maxwell-scalar coupling $\Xi$,
the $\tilde p$-tensor-scalar coupling $\Psi$
and the scalar self-coupling (potential) $V$ are largely unknown,
even in string/M theory \cite{Damour:1994zq},
except in the case where they were obtained in the tree-level superstring
approximation
\cite{Gibbons:1987ps}
- but those were too
simple to describe any realistic phenomena because SUSY gets broken at early stages of the evolution
of the Universe.
Thus, our task now is to find the coupling functions without using the supersymmetrical arguments
solely but engaging instead
some other, physically more general, arguments.

The first thing we require from the theory (\ref{eEMD}) is that
it should not forbid  black hole solutions at any physically relevant $D$
(apart from being compatible with inflation and string/M theory, of course).
Indeed, the mathematical absence of black hole solutions would lead to a loss of  protection of a theory from
naked singularities \cite{Penrose:1964wq} and thus to undesirable violations of the
Cosmic Censorship principle \cite{Penrose:1969pc,Wald:1997wa,Brady:1998au}.
This is especially dangerous in our case: if, according to  modern cosmological views, the Universe
began as an extremely hot, dense and compact higher-dimensional object
then the
ubiquitous fluctuations of the spacetime metric which took place at that stage, inevitably led
to the appearance of  spacetime singularities.
As long as it is impossible to set  initial data in a singular point,
the evolution of physical objects inside the spatial domains which are causally-dependent from singularities
is essentially unpredictable.
Then during the
inflationary  era such regions would become extremely large.
Black holes  protect from this
by ``dressing'' the singularities with event horizons
and thus causally disconnecting them from the rest of the Universe.
Besides, there exist some arguments
that the existence of black holes places an upper
bound for the \textit{cosmological constant} parameter responsible for the rate
of expansion of the Universe and thus prevents the latter from hyper-accelerated inflation \cite{Zloshchastiev:2004ny}.

In view of the aforesaid one may wonder how could it be that black holes, having  such a small
size and mass comparing to those of the Universe, can
nevertheless influence its global properties?
One should recall, however,
that the notion of  event horizon is a non-local one -
in general it requires the knowledge of not only  the distribution of matter in space
but also its future evolution, and depends on the fate of Universe \cite{TPD}.
Thus, the existence of event horizons inevitably affects the large-scale structure of
the whole space.
In the case of  scalar-driven inflationary cosmology
black holes act with the aid
of the global scalar field - it is well-known that
the scalar ``no-hair'' theorems forbid the appearance of black holes for a
large set of  scalar potentials, e.g., convex or positive semi-definite
\cite{Bekenstein:1972ny}.
Thus, by far not every inflaton potential, and hence by far not every
inflationary scenario, is compatible
with the existence of regular horizons
(here we call a horizon \textit{regular} if not only the metric but also
other fields do not become singular on it).

Further, in turn black holes need
for a proper description of their own properties (microstates, entropy, etc.)
certain type of M-theory solutions - branes \cite{Strominger:1996sh}.
Therefore,
the mathematical
absence of brane-like solutions in a theory
would cause, apart from  incompatibility with the main idea of M theory,
 serious difficulties with a consistent microscopical description of
such phenomena as
the black hole thermodynamical laws.
Also, speaking more generally, the branes as extended objects are natural generalizations
of the notion of a point particle and thus can serve as a better approximation
to the quantum-mechanical reality: for instance, one
can ask a question to what extent the
entangled collective quantum states (e.g., the Cooper pairs)
which act as a whole can be regarded as (a plain set of) point-like objects
and is there any better effective description of them.

Lastly, to ensure good ultraviolet behaviour we restrict ourselves
to the class of those p-branes which are explicitly related to M theory
and tend
to  supersymmetric BPS states in some limit \cite{Zloshchastiev:2005mv}.
We demand thus  that our model (\ref{eEMD})
should be compatible with such branes at any physically admissible $D$.
We accentuate that this is a necessary but, of course,
not a sufficient condition for a theory to be physically relevant.
Nevertheless, we will see straight away that it significantly decreases the number of allowed coupling functions.

Imposing this requirement for the 0-branes from the above-mentioned class
(we are looking for necessary conditions, therefore,
we can take 0-branes for simplicity), using the corresponding ansatz from
\Rf \cite{Zloshchastiev:2001mw}
and applying the approach of \Rf \cite{Zloshchastiev:2005mv}
assuming $D$ arbitrary,
we obtain that the most general couplings of (\ref{eEMD}) which allow such 0-branes
are those obeying
the following second-order differential equations
(see the Appendix for more details):
\ba
&&
\hat\Xi''(\phi)
+
\frac{4 \eta}{\phi_1}
\hat\Xi'(\phi)
-
\frac{2(\tp-1)}{\tp}
\hat\Xi(\phi)
=0, ~~
\lb{eCE1}
\\&&
\sfv ''(\phi)
-
\frac{\tp\, \text{coth}(\frac{\phi}{\phi_1})- \eta (\tp - 2)}{\phi_1 (\tp -1)/2}
\sfv '(\phi)
+
\frac{2}{\tp}\sfv(\phi)
=0
,~~
\lb{eCE2}
\ea
where
$
\hat\Xi(\phi) \equiv 2 Q^2/\Xi(\phi) + \tp\, ! P^2 \Psi(\phi),
$
with $Q$ and $P$ being, respectively, the electric and magnetic brane charges (up to a numerical
coefficient),
and
\[
\phi_1 \equiv \frac{4 a}{\varsigma \Delta}, \
\eta \equiv 1-\frac{
4 (\tp -1)}{\tp \Delta}, \
\Delta = a^2 + 2 - \frac{2}{\tp},
\]
$a$ is some constant parameter related to p-branes \cite{Stelle:1998xg}
(here we assume it free as well),
and
$\varsigma$ is $+1$ for elementary (electric) branes and $-1$
for solitonic (magnetic) ones.
The fact that the gauge-scalar couplings $\Xi(\phi)$ and $\Psi(\phi)$
appear in \Eq (\ref{eCE1}) only as the combination $\hat\Xi(\phi)$
is  explained by the existence of electric-magnetic duality between the 1-form ${\cal A}$ (such that $F = d {\cal A}$)
coupled to the worldline of 0-brane and the
$(\tp-1)$-form ${\cal A}_{(\tp-1)}$ (such that $F_{(\tp)} = d {\cal A}_{(\tp-1)}$).

As long as the solutions of \Eqs (\ref{eCE1}), (\ref{eCE2}) could be, in general, complex-valued functions
with cuts and singularities,
the equations must be supplemented with the restriction for the couplings to be physically admissible for real
values of the inflaton.
Otherwise, the action (\ref{eEMD}) would become ill-defined from the field-theoretical point of view.
We impose thus the following ``boundary'' condition:
\be\lb{realcond}
\Xi(\phi), \
\Psi(\phi), \
\sfv (\phi) \subset \text{``physical''} \ \ \text{for all} \ \ \phi \in \Re\text{e},
\ee
where ``physical'' means  a set of at least real, single-valued and regular in a finite real domain functions of $\phi$ -
as the couplings should be.
At first sight, this condition looks too weak to give any physically interesting restrictions.
However, it turns out  that in some cases
it is sufficient to pose a well-defined \textit{eigen}problem \cite{Zloshchastiev:2005mv}.

Further, \Eq (\ref{eCE1}) can be easily solved,
\be
\hat\Xi (\phi) =
\s_1
e^{
\frac{2 \varsigma (\tp -1) }{a \tp}  \phi
}
+
\s_2
e^{- \varsigma a  \phi}
,
\label{eXi-G}
\ee
where $\sigma_{i}$'s are arbitrary integration constants,
and it is of no interest to us here, as \Eq (\ref{realcond}) can not bring any restrictions
for the parameter $\tp$ but only infers that $\sigma$'s must be real-valued.
Thus, it seems that gauge fields, even
those having a
long range, have little or no influence on the process of the Universe's dimensionality formation.

The other coupling equation, for the scalar field potential, looks more promising in this connection.
In fact, it describes the self-interaction of the inflaton which is known to determine
the global dynamics of the Universe, according to the scalar-driven inflationary scenarios.
Could it thus be that \Eq (\ref{eCE2}) determines not only the potential itself but also
the parameter $D$?

Due to the complexity of \Eq (\ref{eCE2}) it seems difficult to  solve exactly the eigenvalue problem
imposed by \Eq (\ref{realcond}) in the general case.
Luckily, one can engage certain physical picture to see things more clearly.
Starting from some cosmological epoch, when the characteristic energy became of the order of the string scale,
the global dynamics of the Universe can be effectively described by the action (\ref{eEMD}) at $D>4$
in the supergravity (SUGRA) approximation \cite{VanNieuwenhuizen:1981ae}.
At that stage the existence of p-branes did not imply any restrictions for $D$ as
the high symmetry
ensured that the model (\ref{eEMD}) has p-brane solutions at any  $D$ provided its couplings
are compatible with supergravity  \cite{Gibbons:1987ps,Stelle:1998xg},
i. e., the  eigenvalues of $D$
have a ``continuous'' spectrum - in a sense that they take integer values (restricted by $D \leq 11$
so as to exclude the appearance of an infinite number of
fields with  spins higher than $2$) but otherwise arbitrary.
However, as the Universe expands and cools,
SUSY gets dynamically broken such that in \Eq (\ref{eEMD}) the SUSY-compatible
couplings
become physically inadmissible (in particular, $V(\phi)$ cannot be set to zero
anymore, otherwise inflation cannot start).
Therefore,
one should go on
computing the general couplings directly from \Eqs (\ref{eCE1}) and (\ref{eCE2}).
Those, of course, will include SUGRA couplings as a special case.

All this can be quantitatively visualized if one clocks the  history of the
early Universe
by means of the cosmological scalar field's scale (of course,
starting from the epoch when this field has already appeared).
Then one could distinguish the following three regimes of \Eq (\ref{eCE2})  and, hence, of $V(\phi)$:

\textit{Top high-energy regime}.
In the pre- and early-inflationary Universe nothing could prevent the
cosmological scalar  from having
large initial magnitude.
This regime thus corresponds to large values of $\phi$ -
such that the cotangent's absolute value
in \Eq (\ref{eCE2}) approaches one.
Actually, ``large'' is a way too strong word here: the cotangent's magnitude
approaches unity exponentially rapidly so that already at $|\phi/\phi_1| \geq 2$ (in Planck units)
it differs from
one by less than $4 \%$.
The general solution of \Eq (\ref{eCE2}) in this regime is a linear
combination of two exponents and no restrictions for $D$ arise from \Eq (\ref{realcond}).
Thus, in this regime the dimensionality parameter is essentially free
that confirms aforesaid.
Notice that the full supersymmetry can be restored because $V=0$ is also a solution of
this equation.

\textit{Bottom high-energy regime}.
As the cosmological time passes, the inflaton's energy density gets diluted by inflation
so
the scalar starts decreasing its magnitude towards its vacuum value
and eventually at some point $|\phi/\phi_1|$ becomes approximately one
in Planck units,
$|\phi/\phi_1| \approx (M_{\text{Planck}})^{\tp/2}$.
In this regime one should consider \Eq (\ref{eCE2}) in full
and study the \textit{eigen}problem by imposing the condition (\ref{realcond}).
The dimensionality of spacetime is still above four
but perhaps some values of $D$ are already forbidden.

\textit{Low-energy regime}.
The Universe continues  expanding
so that
the scalar gradually approaches its recent value,
$\phi \to \phi_{\text{now}}$, which is very small:
$\phi_{\text{now}} \ll (M_{\text{Planck}})^{\tp/2}$.
Therefore, in this regime
the cotangent predominates the constant $\eta$-term in \Eq (\ref{eCE2})
so that
the general solution can be expressed in terms of
the associated Legendre functions \cite{as}:
\be
\sfv (\phi)_{|\phi/\phi_1| \ll 1} =
(z^2 - 1)^{\frac{\tp}{2(1-\tp)}}
\left(
C_1 P_\nu^\mu (z)
+
C_2 Q_\nu^\mu (z)
\right),
\ee
where
$z = \coth (\phi/\phi_1)$, the $C_i$'s are arbitrary constants
(complex-valued, in general), and
\be
\nu \equiv \frac{1}{1 - \tp} - 2, \
\mu \equiv \frac{1}{\tp -1}
\sqrt{
\tp^2 -
\frac{32 \, a^2 (\tp - 1)^2}{\tp \Delta^2}
}
. \lb{definmu}
\ee
Further,  the general solution for $\sfv (\phi)_{|\phi/\phi_1| \ll 1}$ is
a complex-valued function which has a cut
along the real axis for small values of $|\phi/\phi_1|$ (that correspond to large $|z|$).
Therefore, using App. A of \Rf \cite{Zloshchastiev:2005mv} one can show
that the  condition (\ref{realcond}) can be satisfied at arbitrary $C_i$
(by their proper redefinition) if and only if
$\mu$ and $\nu$  both take integer values:
\ba
&&
\nu = n_\nu = 0, \, \pm 1, \, \pm 2,\,  ... \, , \ \
\\&&
\mu
=
n_\mu = 0, \, \pm 1, \, \pm 2,\, ... \, .
\lb{spectr2}
\ea
Using the first of these spectral formulae
we can draw  Table \ref{tab:mua},
from which one can see that the maximal allowed eigenvalue of $D$ is $4$.
In that case the tensor $F_{(\tp)}$ becomes
of a second rank,
hence, we arrive at the model previously studied in \Rfs \cite{Zloshchastiev:2004ny,Zloshchastiev:2005mv}
whereas some of
the above-mentioned 0-branes become 4D black holes with regular scalar and gauge hair.
At that, when the gauge field is off, the regular horizon survives only if the parameter $a$ obeys the
second of the spectral formulae above (incidentally, \Eq (\ref{spectr2}) brings us to the relation
noticed by Hull and Townsend
when dealing with extremal black holes in 4D string compactifications \cite{Hull:1994ys}).
Notice also that in 4D both spectral formulae become exact,
as one can see from \Eq (\ref{eCE2}).

\begin{table}[t]
    \begin{tabular}{cccc}
      \hline \hline
      $~~n_\nu$ ~ & $D-2$ &~~~  $n_\nu$~ & $D-2$\\
      \hline
      $0$ & $1/2$ &  & \\
      $1$ & $2/3$ &  $-1$ & $0$ \\
      $2$ & $3/4$ &  $-2$ & $\pm\infty$ \\
      $3$ & $4/5$ &  $-3$ & $2$ \\
      $4$ & $5/6$ &  $-4$ & $3/2$ \\
      $5$ & $6/7$ &  $-5$ & $4/3$ \\
      $\vdots$ & $\vdots$ &  $\vdots$ & $\vdots$ \\
      $\infty$ & $1$ &  $- \infty$ & $1$ \\
      \hline \hline
    \end{tabular}
    \caption{Correspondence between the spectra of $\nu$ and $\tp$.}
    \label{tab:mua}
\end{table}

Thus,
when the Universe is hot, dense and compact, all the
modes of the scalar potential, described by eigenfunctions of $V(\phi)$,
are equally permitted.
But as the Universe evolves in time, i.e., expands and cools,
the potential occupies one of the few  (eigen)modes which are allowed by
above-mentioned physics.
Other modes become dynamically unstable in the low-energy regime.
At that, the transition from the ``continuous'' (i.e., integer but otherwise arbitrary)
spectrum of $D$ to $D=4$ happens not continuously but by jumps,
due to the discrete nature of $D$.

One can find some similarity between this scenario and the
recombination
of an
electron and a stand-alone hydrogen ion.
When the electron has sufficiently high energy (above the ionization threshold),
its energy spectrum is essentially continuous.
As the energy decreases, e.g., due to the spontaneous emission
of photons, the electron gets captured by the ion and its energy enters a discrete  spectrum.
At that, we know that no description of such binding is possible in terms of classical orbits and trajectories -
instead, one should think in terms of quantum superpositions and transition probabilities.
One can not exclude the possibility that somewhere a similar uncertainty arises in the very early Universe:
there may exist no fully deterministic description of the transition from higher dimensions to 4D
so that the only essential information about this process that one can have in principle is the set of
the allowed coupling (eigen)functions, such as
$
V(\phi)$.
However, the analogy of the above-mentioned quantum-mechanical phenomenon with
our case should be treated with great care because it is not perfect.
In particular,
in the quantum-mechanical case one has a linear superposition of eigenfunctions
which correspond to different eigenvalues of energy whereas
here we have different regimes of the same function, $V(\phi)$,
at different values of its argument.
Thus, the interpretation of our \textit{eigen}problem  is largely an open question so far.
It would be interesting to study it
in full, i.e., when the inflaton potential stays in the above-mentioned ``bottom high-energy'' regime.
This should provide more details about how the transition from larger values of
$D$ proceeds to $D=4$.
Besides, if one has the exact solutions of \Eqs (\ref{eCE1}), (\ref{eCE2}) in hand then using the
method described in \Rf \cite{Zloshchastiev:2001mw}  one can immediately
obtain  exact generalized 0-brane solutions which
are higher-dimensional generalizations of black holes.

To conclude,
we have determined the self-coupling function (potential) of a cosmological scalar field
which would break supersymmetry in such a way as to
be compatible with  inflationary cosmology
and yet preserve some universal features of the low-energy M-theory's solutions.
Assuming the number of spacetime dimensions
to be a free parameter,
we considered the eigenvalue problem for it.
It is shown that in the low-energy regime of
small values of the cosmological scalar field
$D=4$ arises
as a largest allowed eigenvalue.
Hopefully, knowledge of the behaviour of the inflaton potential at different
$D$ will advance also our
understanding of other phenomena which took place in the early Universe.

\begin{acknowledgments}
I acknowledge the fruitful discussion about \Eq (\ref{realcond}) with Hernando Quevedo and Daniel Sudarsky
which greatly helped to clarify this point.
\end{acknowledgments}


\appendix*
\section{Derivation of the coupling equations}

The general 0-brane ansatz can be written in the following form.
For metric we assume
\be
\drm s^2 = - e^{U(r)} \drm t^2 + e^{-U(r)} \drm r^2 +
e^{A(r)} \drm \Omega^2_{(\tp)},                         \lb{z-metr}
\ee
where  $\drm \Omega^2_{(\tp)}$ is a metric of a $\tp$-dimensional $SO$-symmetrical transverse space,
and $r$ is the radial coordinate therein.
The Maxwell and $\tp$-form fields are assumed being in the form
\be
F = Q \frac{e^{- \tp A/2}}{\Xi (\phi)}  \drm t \wedge \drm r, \ \
F_{(\tp) M...N} = P e^{- \tp A/2} \varepsilon_{M...N}
\ee
in an orthonormal frame, and $\phi$ is just a function of $r$.
From the corresponding equations of motion one can derive the so-called
class equation \cite{Zloshchastiev:2001mw}:
\bw
\ba
&&
\frac{H_{,\phi}}{\ap_{,\phi}}
+
\left(
\frac{1}{\tp \ap_{,\phi}^2} + \frac{\tp-1}{2}
\right)
H
+
\frac{e^\ap}{\tp}
\left(
      \Lambda + e^{- \tp \ap} \hat\Xi
\right)
+
\tp
=1,                         \lb{z-g1a}
\ea
where
\be
H \equiv
\frac{1}{\tp (1/\ap_{,\phi})_{,\phi}}
\biggl[
\tp (\tp-1) +
e^\ap
\left(
      \Lambda + \frac{\tp }{2} \Lambda_{,\phi} \ap_{,\phi}
\right)
+
e^{-(\tp-1) \ap}
\left(
      \hat\Xi + \frac{\tp }{2} \hat\Xi_{,\phi} \ap_{,\phi}
\right)
\biggr]
,
\ee
\ew
where the subscript ``$,\phi$'' stands for the derivative
with respect to $\phi$.

Now, suppose we have fixed  the function $\ap (\phi)$.
Then the class equation turns into a joint linear second-order ODE
with unknown functions $\Lambda (\phi)$ and $\hat\Xi (\phi)$.
This ODE thus is, in fact,
the constraint for $\Lambda$ and $\hat\Xi$ which ensures
the internal consistency of the theory.
Therefore, with any given $\ap (\phi)$
one can associate the appropriate
class of integrability given by a self-consistent
$\{\Lambda,\,\hat\Xi\}$ pair.
Thus, the space of all possible coupling functions becomes
``inhomogeneous'' as it can be divided according to the class structure.
Theories belonging to a class are equivalent in a sense
that their appropriate ground-state solutions possess the same dependence $\ap (\phi)$.

The next step is to find the most general class of equivalence which contains the static p-brane
solutions related to M theory and possessing a SUSY limit, i.e., reducing to BPS states
at some values of their parameters \cite{Stelle:1998xg}.
The latter will also ensure that quantum corrections are taken into account
because the BPS states which define the class are protected from them
and thus keep the class unchanged.
The characteristic function $A(\phi)$ for this class
is given by \Eq (7) from \Rf \cite{Zloshchastiev:2005mv},
and for a 0-brane reduces to
\be
A (\phi) =
\frac{2 \varsigma }
   {a \tp } \phi
+
  \frac{2}{{\tp-1}} \ln \k
-
  \frac{2}
     {{\tp-1}}
     \ln (
       e^{\frac{\Delta \,\varsigma \,\phi }{2\,a}} -1)
,
\lb{eBC}
\ee
where
$\k$ is the integration constant which sets the brane's mass scale.
Then we
substitute it into the class equation above and obtain:
\bw
\be
\hat\Xi''(\phi)
+
\frac{4 \eta}{\phi_1}
\hat\Xi'(\phi)
-
\frac{2(\tp-1)}{\tp}
\hat\Xi(\phi)
+ \kappa^{\frac{2 \tp}{\tp -1}}
f(\phi)
\left[
\sfv ''(\phi)
-
\frac{\tp\, \text{coth}(\frac{\phi}{\phi_1})- \eta (\tp - 2)}{\phi_1 (\tp -1)/2}
\sfv '(\phi)
+
\frac{2}{\tp}\sfv(\phi)
\right]
=0
,~~
\lb{eCE}
\ee
\ew
where $f(\phi)$ is some function whose explicit form
is not important here.
Finally, as long as an action functional and hence coupling functions must be independent from the mass scale parameter $\kappa$
(see \Rf \cite{Zloshchastiev:2005mv}, Sec. III) we can separate the coupling equations.

\def\AnP{Ann. Phys.}
\def\APP{Acta Phys. Polon.}
\def\CJP{Czech. J. Phys.}
\def\CMPh{Commun. Math. Phys.}
\def\CQG {Class. Quantum Grav.}
\def\EPL  {Europhys. Lett.}
\def\IJMP  {Int. J. Mod. Phys.}
\def\JMP{J. Math. Phys.}
\def\JPh{J. Phys.}
\def\FP{Fortschr. Phys.}
\def\GRG {Gen. Relativ. Gravit.}
\def\GC {Gravit. Cosmol.}
\def\LMPh {Lett. Math. Phys.}
\def\MPL  {Mod. Phys. Lett.}
\def\Nat {Nature}
\def\NCim {Nuovo Cimento}
\def\NPh  {Nucl. Phys.}
\def\PhE  {Phys.Essays}
\def\PhL  {Phys. Lett.}
\def\PhR  {Phys. Rev.}
\def\PhRL {Phys. Rev. Lett.}
\def\PhRp {Phys. Rept.}
\def\RMP  {Rev. Mod. Phys.}
\def\TMF {Teor. Mat. Fiz.}
\def\prp {report}
\def\Prp {Report}

\def\jn#1#2#3#4#5{{#1}{#2} {\bf #3}, {#4} {(#5)}} 

\def\boo#1#2#3#4#5{{\it #1} ({#2}, {#3}, {#4}){#5}}



\end{document}